\documentclass[journal,twoside]{IEEEtran}
\usepackage[T1]{fontenc}
\usepackage[utf8]{inputenc}
\usepackage[table,dvipsnames,svgnames,x11names]{xcolor}
\usepackage{cite}

\usepackage{enumitem}
\usepackage{array}
\usepackage{makecell}

\newcolumntype{P}[1]{>{\centering\arraybackslash}m{#1}}
\usepackage{amsmath,amssymb,amsfonts}
\usepackage{algorithmic}
\usepackage{graphicx}
\usepackage[mathlines,switch]{lineno}
\usepackage{textcomp}
\usepackage{newunicodechar}
\newunicodechar{₃}{$_3$}
\usepackage{wrapfig,colortbl}
\usepackage{blindtext}
\usepackage{lipsum}
\usepackage[final]{changes}
\usepackage{xcolor}

\setaddedmarkup{\textcolor{blue}{#1}}
\setdeletedmarkup{\textcolor{red}{\sout{#1}}}
\usepackage{TUSON}
\colorlet{nblue}{black}
\colorlet{mblue}{black}

\hypersetup{hidelinks}

\makeatletter
\renewenvironment{abstract}
  {\global\setbox0=\vbox\bgroup\oabstract}
  {\oendabstract\egroup\box0}
\makeatother

\def\BibTeX{{\rm B\kern-.05em{\sc i\kern-.025em b}\kern-.08em
    T\kern-.1667em\lower.7ex\hbox{E}\kern-.125emX}}

\title{In-Plane Q Anisotropy of Higher-Order XBARs \\in 128$^\circ$Y-Cut Lithium Niobate}
\author{Byeongjin Kim, Ian Anderson, and Ruochen Lu
\thanks{This work was supported by the National Science Foundation (NSF) under CAREER Award No.\ 2339731, and a graduate fellowship from the Ministry of National Defense of the Republic of Korea.}
\thanks{Byeongjin Kim, Ian Anderson, and Ruochen Lu are with the Electrical and Computer Engineering Department, The University of Texas at Austin (email: c18263@utexas.edu).}}

\IEEEaftertitletext{\GA{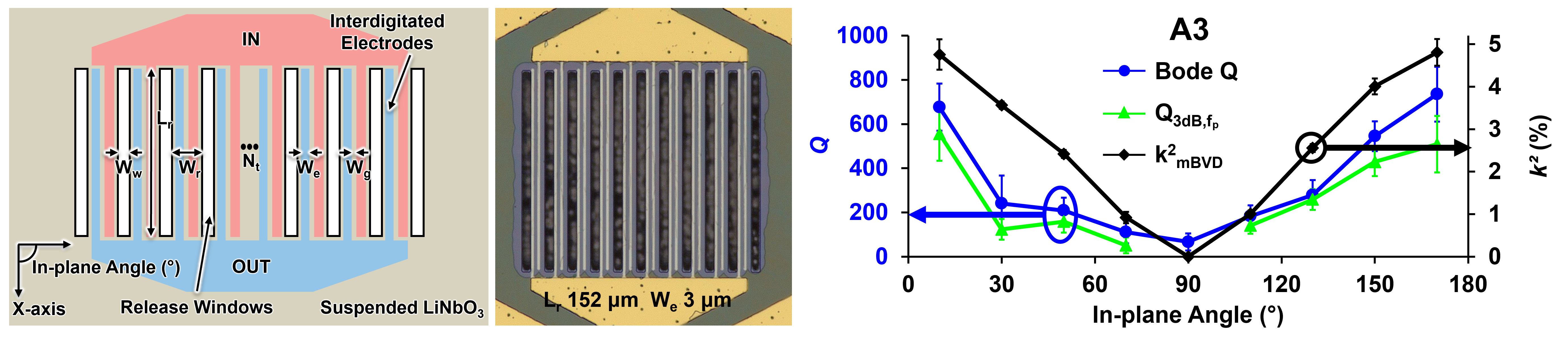} 
\Abstract{\begin{abstract}
128$^\circ$Y-cut lithium niobate (LN) laterally field-excited higher-order antisymmetric bulk acoustic resonators (XBARs) have attracted interest for high-frequency acoustic devices thanks to their high electromechanical coupling coefficient ($k^2$), high quality factor ($Q$) from low metal coverage ratio, and thickness-defined resonant frequency. So far, the in-plane orientation of these resonators is commonly chosen to maximize $k^2$, thereby maximizing bandwidth. More recently, in-plane-rotated XBARs in 128$^\circ$Y-cut LN have been built to provide greater design flexibility in filter synthesis. However, the in-plane anisotropy of $Q$ has been far less explored. This leaves an important gap in understanding whether the propagation direction that determines $k^2$ also affects $Q$. In this work, we investigate the anisotropic $Q$ of higher-order antisymmetric modes (namely, A$_3$, A$_5$, and A$_7$) in 500-nm-thick 128$^\circ$Y-cut LN on Si. By characterizing resonator performance in various in-plane orientations, we observe that both Bode $Q$ and $Q_{\mathrm{3dB}, f_p}$ show minimum values at 90$^\circ$ to the material x-axis and maximum values around 0$^\circ$, following a trend similar to $k^2$. The A$_3$, A$_5$, and A$_7$ modes around 10.4, 17, and 24 GHz exhibit averaged Bode $Q$/$Q_{\mathrm{3dB}, f_p}$ values of 735/546, 204/149, and 59/37, respectively. At 90$^\circ$, the average Bode $Q$ values are reduced to 66, 9, and 12. Finite element analysis (FEA) results suggest that the orientation-dependent degradation of $Q$ near 90$^\circ$ is associated with stronger transverse displacement near the inactive and anchor regions, resulting in enhanced energy leakage. These results reveal an orientation-dependent loss pathway in 128$^\circ$Y-cut LN XBARs and provide design guidance for jointly optimizing $k^2$ and $Q$.
\hblne
\end{abstract}}
\vspace{1\baselineskip}\vspace*{-1pt}
}

\begin{document}

\maketitle

\Highlights[1]{Anisotropic $Q$ of higher-order antisymmetric modes (A$_3$/A$_5$/A$_7$) measured in 128$^\circ$Y-cut LN on Si}

\Highlights[2]{Averaged Bode $Q$ of 735/204/59 (A$_3$/A$_5$/A$_7$) follows $k^2$, dropping sharply to 66/9/12 at 90$^\circ$}

\Highlights[3]{FEA suggests that reduced $Q$ near 90$^\circ$ is consistent with enhanced transverse motion near the anchors}

\Keywords{128$^\circ$Y-cut lithium niobate, anchor loss, anisotropy, higher-order antisymmetric mode, laterally excited bulk acoustic resonator (XBAR), quality factor}

\PrintHighlights

\section{Introduction}
\label{sec:introduction}

\IEEEPARstart{A}{coustic} resonators are essential components in modern electronic systems, including radio-frequency (RF) front-end filters \cite{ruppelAcousticWaveFilter2017}, oscillators \cite{abdolvandThinfilmPiezoelectriconsiliconResonators2008}, sensors \cite{yantchevThinFilmLamb2013}, and emerging quantum acoustic platforms \cite{wollackQuantumStatePreparation2022}. Their compact footprint, frequency scalability, and efficient electromechanical transduction make them attractive for highly integrated signal-processing systems \cite{rubySnapshotTimeFuture2015,gongMicrowaveAcousticDevices2021}. Among various piezoelectric materials, lithium niobate (LN) has emerged as a promising platform for high-performance acoustic resonators due to its strong piezoelectricity and large electromechanical coupling coefficient ($k^2$) \cite{luRecentAdvancesHighperformance2025c,luRFAcousticMicrosystems2021a,pillaiPiezoelectricMEMSResonators2021}. In particular, 128$^\circ$Y-cut LN has been widely explored for laterally excited bulk acoustic resonators (XBARs), where strong coupling has enabled high-frequency acoustic devices \cite{liA1ModeLambWave2024,kramerAcousticResonators1002025a,barrera50GHzPiezoelectric2026}.

As wireless communication systems continue to evolve toward higher frequencies and wider bandwidths, both $k^2$ and quality factor ($Q$) become critical performance metrics \cite{yang1060GHzElectromechanicalResonators2020a,kramerCryogenicEnhancement502026,tabrizianEffectPhononInteractions2009,hagelauerMicrowaveAcousticFilters2023a}. While $k^2$ determines the energy transduction efficiency and achievable bandwidth, $Q$ governs acoustic energy dissipation and directly affects insertion loss, frequency selectivity, and overall resonator efficiency \cite{anusornPracticalDemonstrationsFR3Band2025,luRFAcousticMicrosystems2021a}. Therefore, understanding how both $k^2$ and $Q$ vary with device design and material orientation is essential for optimizing LN acoustic resonators \cite{gongDesignAnalysisLithium2013,popInvestigationElectromechanicalCoupling2018,wuKBandLiNbO3A32025,faizanOptimizationInactiveRegions2021}.

Because LN is strongly anisotropic, the in-plane propagation orientation has been widely used as a key design parameter for improving resonator performance. In 128$^\circ$Y-cut LN, previous studies have primarily utilized this anisotropy to identify orientations that maximize $k^2$, particularly for high-coupling XBARs \cite{kramerGeneralizedAcousticFramework2025a,liA1ModeLambWave2024}. However, optimizing $k^2$ alone does not necessarily guarantee high $Q$.

Recent studies have begun to show that acoustic dissipation in LN resonators can also depend strongly on propagation orientation. For example, anisotropic $Q$ behavior has been reported in X-cut LN S$_0$ resonators \cite{popInvestigationElectromechanicalCoupling2018} and in 36$^\circ$Y-cut LN SH$_0$ resonators \cite{liuHighFigureofMerit36YCut2025}. In these platforms, the resonance quality factor was shown to vary with in-plane propagation angle, indicating that acoustic loss is also governed by crystal anisotropy. These studies suggest that maximizing $k^2$ alone is insufficient for realizing optimal resonator performance and that orientation-dependent loss mechanisms must also be considered. Nevertheless, a systematic investigation of anisotropic $Q$ in 128$^\circ$Y-cut LN XBARs remains limited, despite the importance of this cut for high-frequency, high-coupling acoustic devices.

In this work, we experimentally investigate the anisotropic $Q$ of higher-order antisymmetric modes in 500-nm-thick 128$^\circ$Y-cut LN on Si. By characterizing A$_3$, A$_5$, and A$_7$ modes as a function of in-plane propagation orientation, we observe that both Bode $Q$ and 3dB$Q$ at the parallel resonant frequency ($Q_{\mathrm{3dB}, f_p}$) follow a trend similar to $k^2$, with minimum values at 90$^\circ$ and maximum values at 0$^\circ$ and 180$^\circ$. The A$_3$, A$_5$, and A$_7$ modes around 10.4, 17, and 24 GHz exhibit averaged Bode $Q$/$Q_{\mathrm{3dB}, f_p}$ values of 735/546, 204/149, and 59/37, respectively. At 90$^\circ$, the average Bode $Q$ values are reduced to 66, 9, and 12 for the A$_3$, A$_5$, and A$_7$ modes, while $Q_{\mathrm{3dB}, f_p}$ cannot be reliably extracted because of the weak resonance response. FEA results further indicate that unintended transverse displacement is strongest at 90$^\circ$, particularly near the anchors, leading to increased acoustic energy leakage and reduced $Q$. These results suggest that the orientation-dependent degradation of $Q$ in 128$^\circ$Y-cut LN XBARs is strongly influenced by transverse mode deformation and anchor-region leakage, providing design guidelines for simultaneously optimizing $k^2$ and $Q$.

\section{Design and Simulation}

\begin{figure}[t!]
\centerline{\includegraphics[width=\columnwidth]{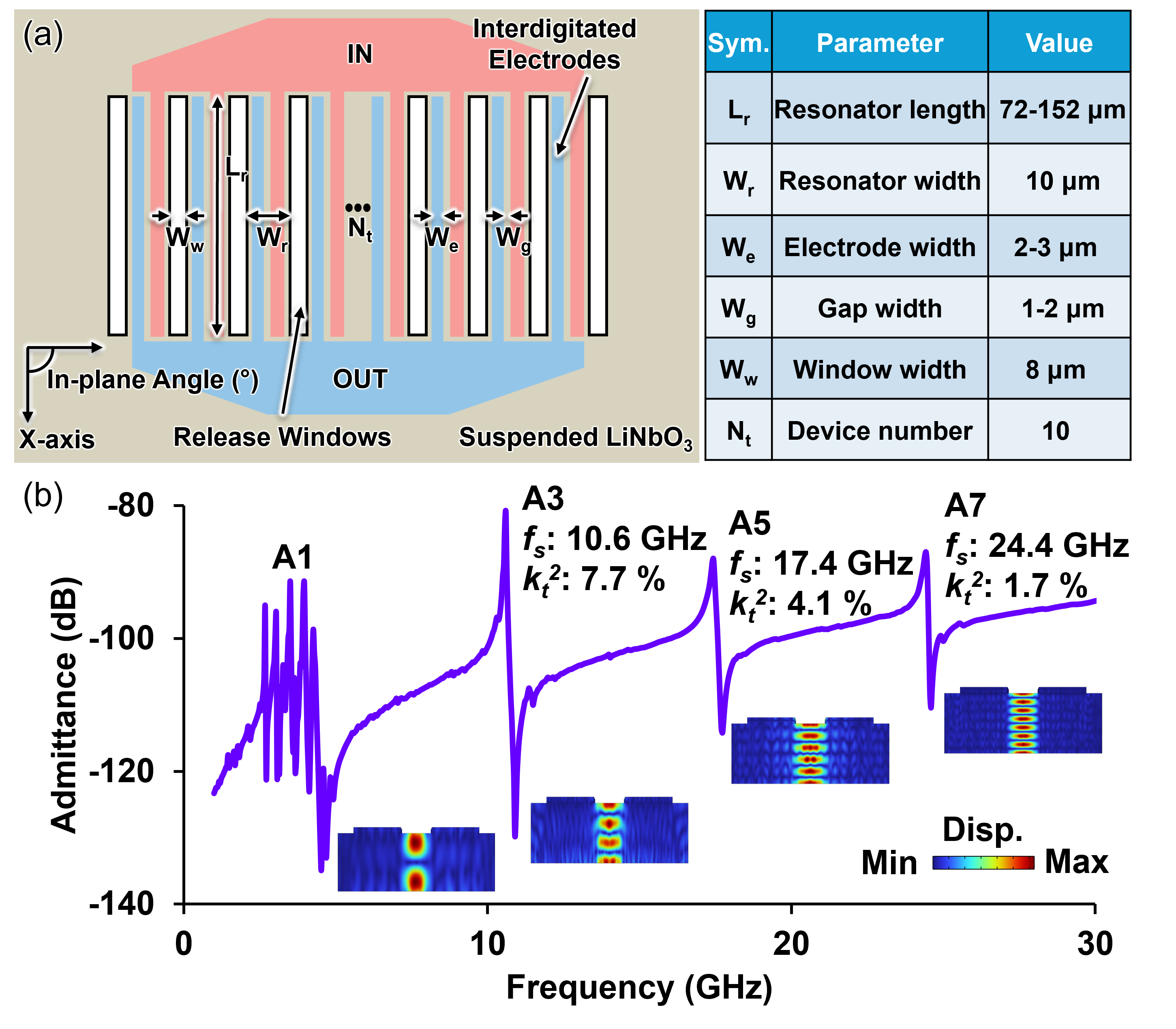}}
\caption{(a) Schematic of the spurious-mode-suppressed 128$^\circ$Y-cut LN resonator design with key dimensions summarized in the table. (b) Simulated wideband admittance response of the resonator array with representative displacement mode shapes.}
\label{fig1}
\end{figure}

To investigate the orientation-dependent behavior of 128$^\circ$Y-cut LN resonators, the device geometry was designed as shown in Fig.~\ref{fig1}(a). The resonator consists of interdigitated transducers (IDTs) patterned on a suspended 500-nm-thick LN resonant body with release windows defined around the active region. A series of MEMS resonators are arranged with respect to the material $x$-axis, and the in-plane angle is defined as the counterclockwise rotation angle from the $x$-axis. The key design parameters, including resonator length ($L_r$), resonator width ($W_r$), electrode width ($W_e$), gap width ($W_g$), window width ($W_w$), and device number ($N_t$), are summarized in the table in Fig.~\ref{fig1}(a). The resonator array was designed to suppress spurious modes and obtain cleaner resonance responses from the intended antisymmetric modes, enabling a more reliable evaluation of anisotropic $Q$ \cite{luExploitingParallelismResonators2018,songArrayingSH0Lithium2016}. Fig.~\ref{fig1}(b) shows the simulated wideband admittance response of the resonator array together with representative displacement mode shapes. Clear resonances are observed for the A$_3$, A$_5$, and A$_7$ modes at approximately 10.6, 17.4, and 24.4 GHz, respectively, each showing a well-defined thickness-shear displacement profile. In contrast, the A$_1$ mode exhibits multiple spurious responses around the main resonance. This is likely because the duty cycle, defined as the ratio of the electrode width to the IDT period, and the LN thickness are not optimized to match the dispersion between the metalized and non-metalized regions of the A$_1$ mode in the present stack. The resulting velocity mismatch can perturb the lateral mode profile and introduce multiple higher-order spurious responses around the main resonance \cite{yangHighAntisymmetricMode2020,yangLateralSpuriousMode2021}. Therefore, the A$_1$ mode is excluded from the anisotropic $Q$ analysis, and the following study focuses on the A$_3$, A$_5$, and A$_7$ higher-order antisymmetric modes, which provide clearer modal responses.

\begin{figure}[t!]
\centerline{\includegraphics[width=\columnwidth]{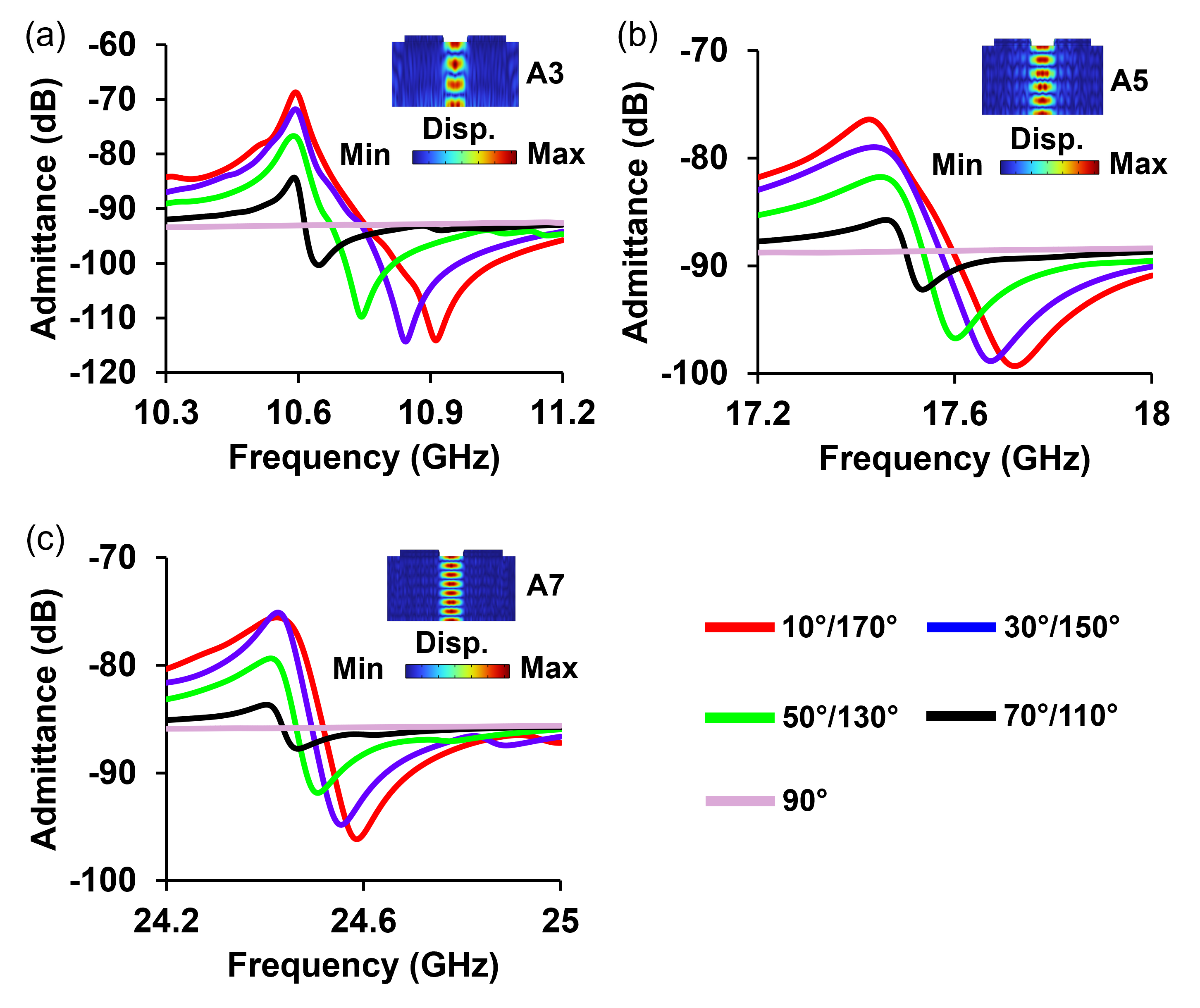}}
\caption{Simulated admittance responses of the 128$^\circ$Y-cut LN resonators at different in-plane angles from 10$^\circ$ to 170$^\circ$: (a) A$_3$ mode; (b) A$_5$ mode; (c) A$_7$ mode. The reduction in $k_t^2$ toward 90$^\circ$ is observed for all three modes.}
\label{fig2}
\end{figure}

The effective $k_t^2$ in spurious-free resonators with high $Q$ can be approximated by:
\begin{equation}
k_t^2 = \frac{\pi^2}{8}\left[\left(\frac{f_p}{f_s}\right)^2 - 1\right],
\label{eq_kt2}
\end{equation}
where $f_s$ and $f_p$ are the simulated series and parallel resonant frequencies. This $f_s$/$f_p$-based $k_t^2$ is equivalent to the $k^2$ value later extracted from the modified Butterworth--Van Dyke (mBVD) model. However, because routing resistance, parasitic capacitance, and other high-frequency parasitics can distort the apparent $f_s$ and $f_p$ spacing, the measured $k^2$ values are extracted using mBVD fitting \cite{naikElectromechanicalCouplingConstant1998,luAccurateExtractionLarge2019,larsonModifiedButterworthVanDyke2000}.

For the higher-order antisymmetric modes in 128$^\circ$Y-cut LN, the excitation is mainly associated with the piezoelectric stress coefficient $e_{15}$, which varies strongly with in-plane propagation angle \cite{kramerGeneralizedAcousticFramework2025a,liA1ModeLambWave2024}. This angular dependence leads to the $k_t^2$ trend shown later in \rev{\mbox{Fig.~\ref{fig8}}}, where $k_t^2$ reaches its minimum at 90$^\circ$ and maximum values at 0$^\circ$ and 180$^\circ$. Consistent with this trend, the simulated admittance responses in Fig.~\ref{fig2} show that, for the A$_3$, A$_5$, and A$_7$ modes, $f_s$ and $f_p$ become closer as the propagation angle approaches 90$^\circ$, indicating reduced $k^2$.

Although the simulations capture the anisotropic coupling trend, the experimentally measured $Q$ is difficult to predict in advance because it depends on multiple loss contributions. The total quality factor is determined by the ratio of stored to dissipated energy and can be expressed as
\begin{equation}
\begin{split}
Q_{\mathrm{total}} 
&= 2\pi \frac{E_{\mathrm{stored}}}{E_{\mathrm{dissipated}}} \\
&= \left(
Q_{\mathrm{anchor}}^{-1}
+Q_{\mathrm{TED}}^{-1}
+Q_{\mathrm{other}}^{-1}
\right)^{-1}.
\end{split}
\label{eq_qtotal}
\end{equation}
where $E_{\mathrm{stored}}$ and $E_{\mathrm{dissipated}}$ are the stored and dissipated energy per cycle, respectively, and $Q_{\mathrm{anchor}}$, $Q_{\mathrm{TED}}$, and $Q_{\mathrm{other}}$ correspond to anchor loss, thermoelastic damping (TED), and other loss mechanisms, respectively \cite{tuDissipationAnalysisMethods2020}. Since these loss contributions depend on material properties, electrode conditions, fabrication variations, and resonator geometry, predefined material quality factors were assigned to LN and the electrodes as practical damping parameters, with $Q_{\mathrm{LN}}=400$ and $Q_{\mathrm{electrode}}=20$, to obtain finite simulated admittance responses. These values are used only to compare relative resonance behavior and mode shapes, while the actual anisotropic $Q$ must be determined experimentally.

\begin{figure}[t!]
\centerline{\includegraphics[width=\columnwidth]{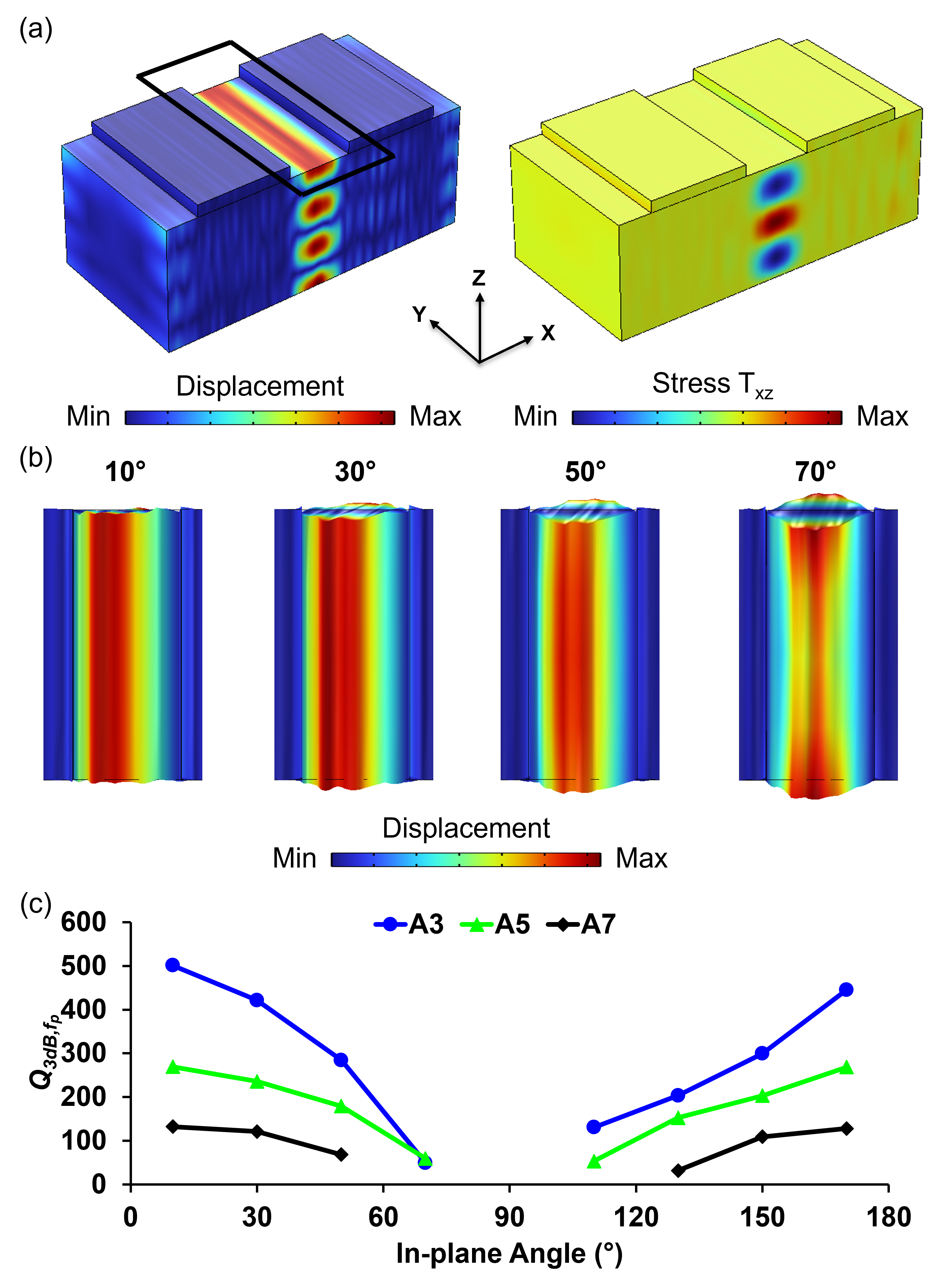}}
\caption{Simulated A$_3$ mode shape and orientation-dependent deformation of the 128$^\circ$Y-cut LN resonator: (a) unit-cell displacement and $T_{xz}$ stress distributions; (b) enlarged top view displacement profiles with deformation at in-plane angles of 10$^\circ$, 30$^\circ$, 50$^\circ$, and 70$^\circ$; (c) simulated $Q_{\mathrm{3dB},f_p}$ of the A$_3$, A$_5$, and A$_7$ modes obtained from a finite one-pair-IDT resonator model at different in-plane angles.}
\label{fig3}
\end{figure}

Using this finite element analysis (FEA) framework, the simulated mode-shape results in Fig.~\ref{fig3} further clarify the orientation-dependent deformation of the A$_3$ mode. Fig.~\ref{fig3}(a) shows the unit-cell displacement and $T_{xz}$ stress distribution, confirming that the intended mode is dominated by thickness-shear deformation. The enlarged top-view displacement profiles in Fig.~\ref{fig3}(b) reveal a clear orientation-dependent deformation behavior. As the in-plane propagation angle increases from 10$^\circ$ to 70$^\circ$ and approaches 90$^\circ$, the unintended transverse displacement component, corresponding to the $y$-direction, becomes increasingly pronounced. The 90$^\circ$ case is not included in Fig.~\ref{fig3}(b) because a well-defined A$_3$ resonance could not be identified at this orientation. Figs.~\mbox{\ref{fig3}}(a) and \mbox{\ref{fig3}}(b) were obtained from a periodic active-region model, which does not include the inactive or anchor regions. Therefore, these profiles are used to identify orientation-dependent transverse deformation, not to directly calculate anchor loss or absolute $Q$. To provide a more quantitative FEM comparison, additional finite one-pair-IDT resonator simulations were performed for the A$_3$, A$_5$, and A$_7$ modes at in-plane angles from 10$^\circ$ to 170$^\circ$ in 20$^\circ$ increments. The simulated $Q_{\mathrm{3dB}, f_p}$ was extracted from the admittance response and is shown in Fig.~\mbox{\ref{fig3}}(c), while the corresponding one-pair-IDT resonator model and anchor-region displacement fields are further examined in Fig.~\mbox{\ref{fig10}}. Data points for which the resonance was too weak to reliably extract the $Q_{\mathrm{3dB}, f_p}$ were omitted. The reduced $Q$ at orientations with larger transverse deformation indicates that this deformation can couple to the inactive and anchor regions in the finite resonator, leading to enhanced acoustic energy leakage.

These simulation results motivate the experimental investigation of resonators with different in-plane propagation orientations. While Fig.~\ref{fig2} shows that $k_t^2$ decreases toward 90$^\circ$, Fig.~\mbox{\ref{fig3}} suggests that the unintended transverse displacement becomes stronger and the simulated $Q_{\mathrm{3dB}, f_p}$ decreases in the same angular range. Therefore, measuring the actual resonator responses across different orientations is necessary to determine whether the anisotropic $Q$ trend is accompanied by the anisotropic coupling trend and to clarify the dominant loss behavior in 128$^\circ$Y-cut LN XBARs.

\section{Fabrication and Measurement}

\begin{figure}[t!]
\centerline{\includegraphics[width=\columnwidth]{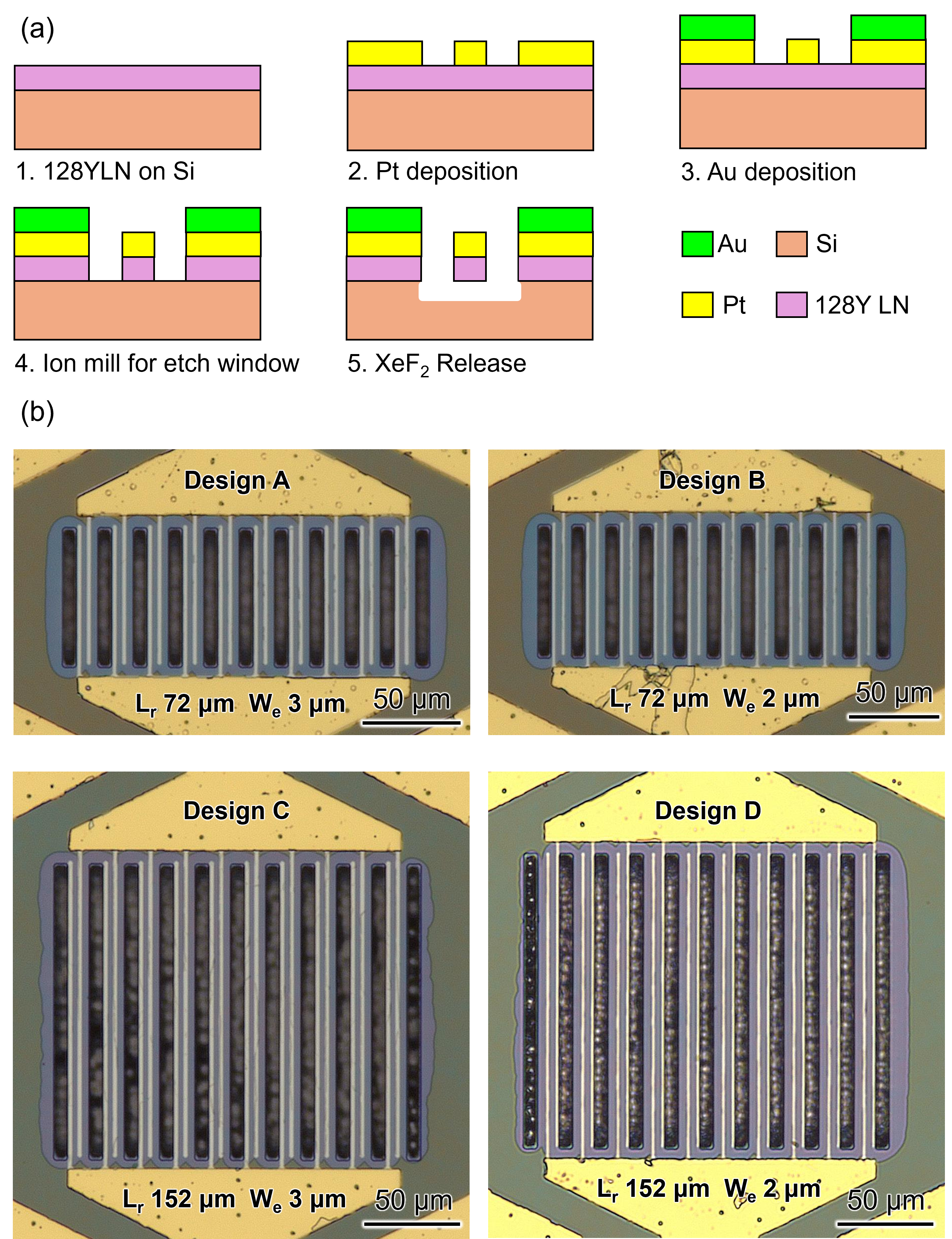}}
\caption{(a) Fabrication flow of the suspended 128$^\circ$Y-cut LN resonators. (b) Optical micrographs of fabricated resonators with different resonator lengths and electrode widths.}
\label{fig4}
\end{figure}

XBAR resonator arrays were fabricated from 500-nm-thick 128$^\circ$Y-cut LN on Si substrates provided by NGK Corporation. Fig.~\ref{fig4}(a) shows the fabrication flow of the suspended resonators. First, fine IDT electrodes were defined by photolithography using an MA8 mask aligner, followed by 50 nm Pt deposition using e-beam evaporation and lift-off in PG remover with sonication. A second photolithography step was then performed to define the buslines and probing pads, followed by 80 nm Au deposition and lift-off. After metal patterning, release windows were defined by photolithography and etched through the LN layer using ion milling. Finally, the underlying Si was removed using XeF$_2$ isotropic etching to form suspended 128$^\circ$Y-cut LN resonators. Fig.~\ref{fig4}(b) shows optical micrographs of the fabricated resonators. Four resonator designs, labeled A--D, were fabricated by varying the resonator length and electrode width. For each design, devices were arranged at different in-plane propagation angles from 10$^\circ$ to 170$^\circ$ in 20$^\circ$ increments, measured with respect to the material $x$-axis. This design set enables statistical comparison of resonators with different dimensions at each propagation orientation, allowing the anisotropic $Q$ trend to be evaluated more reliably.

\begin{figure}[t!]
\centerline{\includegraphics[width=\columnwidth]{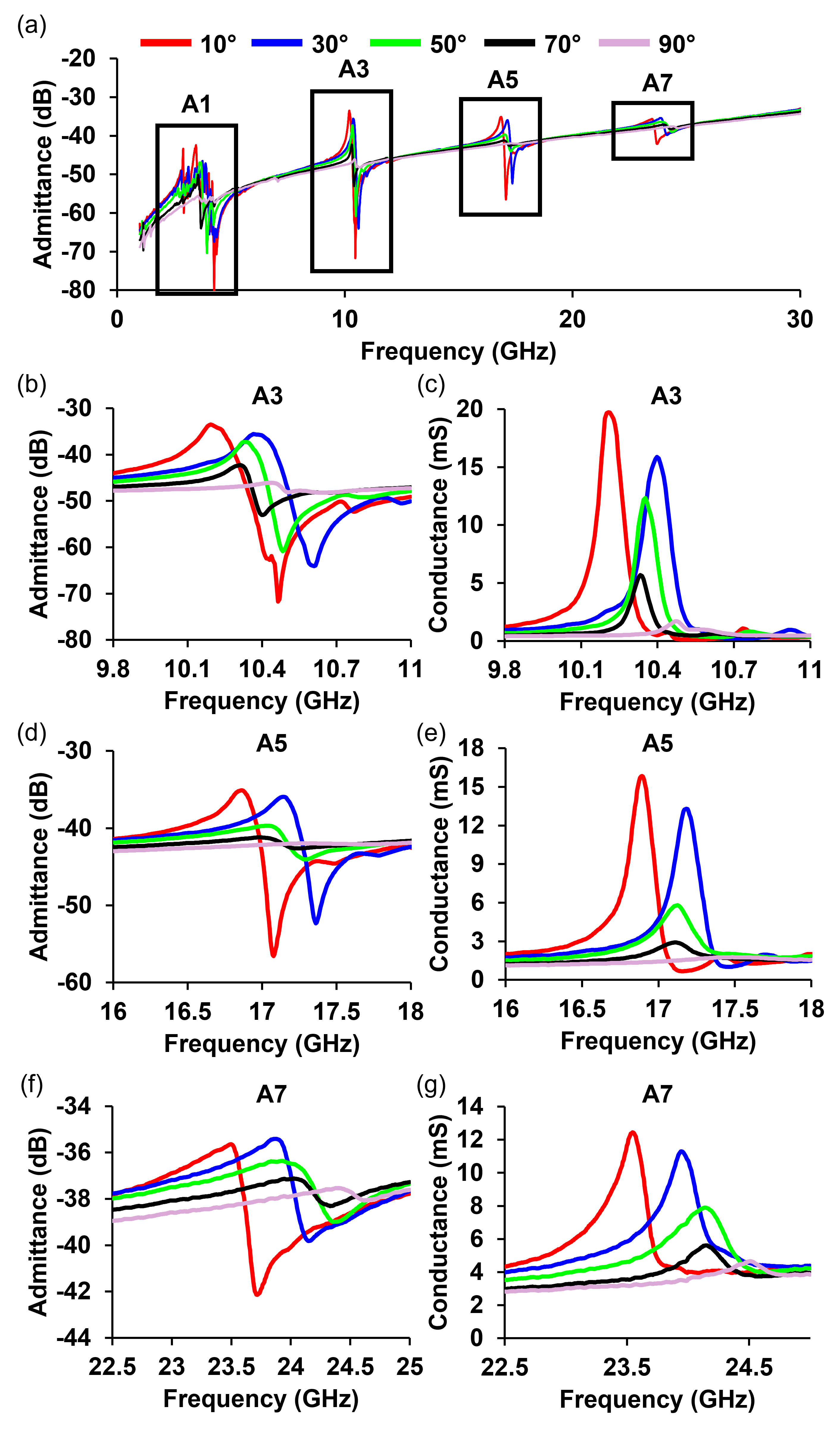}}
\caption{(a) Measured wideband admittance responses of the Design A resonators at different in-plane angles from 10$^\circ$ to 90$^\circ$. Enlarged admittance responses of the (b) A$_3$, (d) A$_5$, and (f) A$_7$ modes. Corresponding conductance responses of the (c) A$_3$, (e) A$_5$, and (g) A$_7$ modes.}
\label{fig5}
\end{figure}

\begin{figure}[t!]
\centerline{\includegraphics[width=\columnwidth]{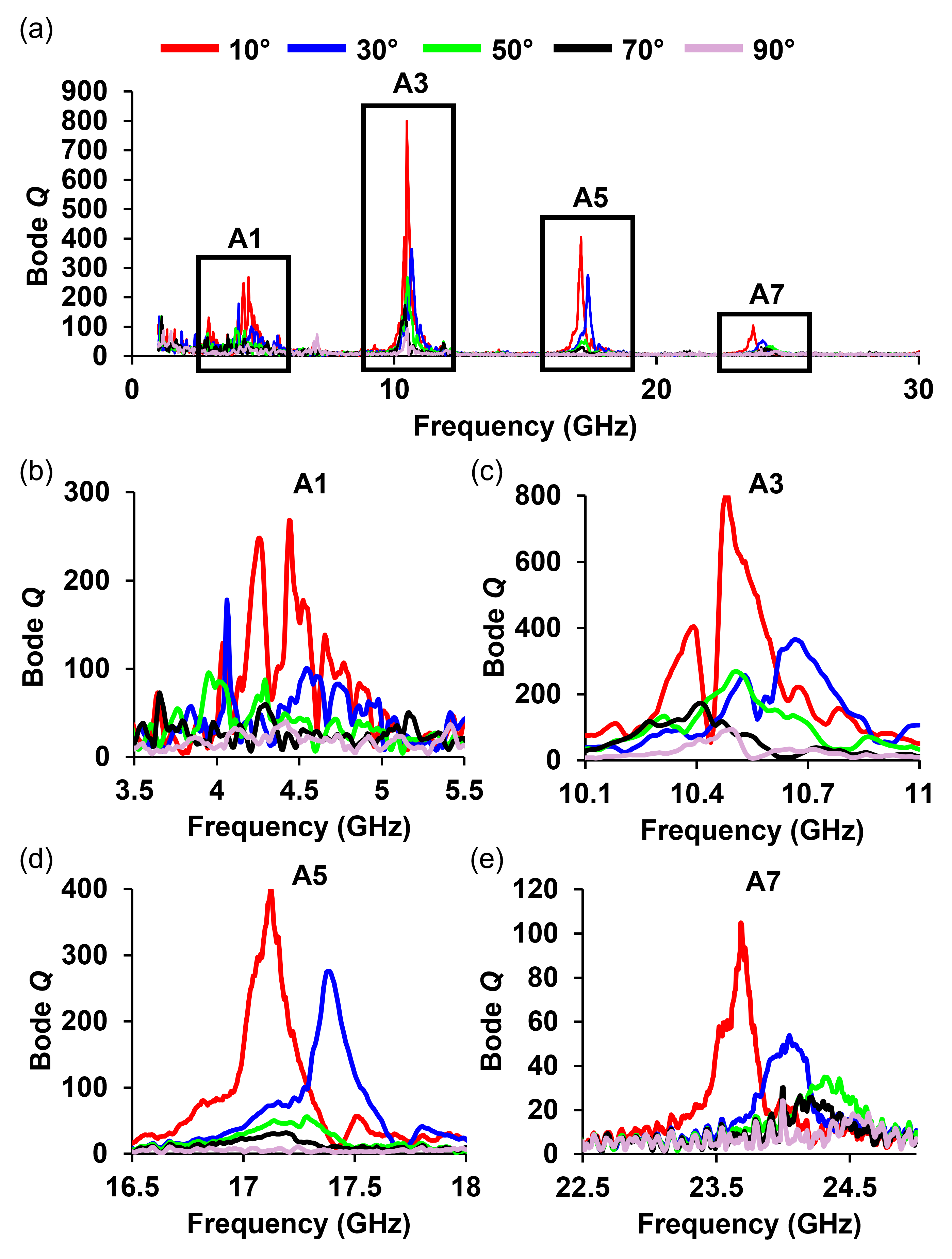}}
\caption{(a) Measured wideband Bode $Q$ responses of the Design A resonators at different in-plane angles from 10$^\circ$ to 90$^\circ$. Enlarged Bode $Q$ responses of the (b) A$_1$, (c) A$_3$, (d) A$_5$, and (e) A$_7$ modes.}
\label{fig6}
\end{figure}

The measured admittance and Bode $Q$ responses shown in Figs.~\mbox{\ref{fig5}} and \mbox{\ref{fig6}} were obtained from Design A, whose complete structural parameters are summarized in Fig.~\mbox{\ref{fig4}}. Therefore, the resonators compared across different in-plane angles in these figures have identical geometry, with $L_r = 72~\mu\mathrm{m}$ and $W_e = 3~\mu\mathrm{m}$. The Bode $Q$ was extracted using the improved $S_{11}$-parameter-based method after converting the measured admittance response to $S_{11}$~\mbox{\cite{jinImprovedFormulaEstimating2021}}. Fig.~\ref{fig5}(a) shows the measured wideband admittance responses of the fabricated resonators at different in-plane angles from 10$^\circ$ to 90$^\circ$. The enlarged measured admittance responses of the A$_3$, A$_5$, and A$_7$ modes around 10.4, 17, and 24~GHz are shown in Fig.~\mbox{\ref{fig5}}(b), (d), and (f), respectively, while the corresponding conductance responses are presented in Fig.~\mbox{\ref{fig5}}(c), (e), and (g). Compared with the admittance, the conductance provides a clearer visualization of the resonance linewidth, peak strength, and possible lossy or spurious responses, particularly at weak-coupling orientations. As the propagation angle approaches 90$^\circ$, the conductance peaks become progressively weaker and broader before eventually becoming indistinguishable from the background response. The increased resonance linewidth observed in the conductance responses is consistent with the reduction in the measured Bode $Q$ shown in Fig.~\mbox{\ref{fig6}}. Small differences between the measured resonance frequencies and the simulated values, as well as device-to-device frequency variations, are mainly attributed to thickness nonuniformity within the same LN-on-Si sample. As expected from the simulated $k_t^2$ trend, $f_s$ and $f_p$ become closer as the propagation angle approaches 90$^\circ$, indicating reduced $k_t^2$. At 90$^\circ$, the resonance response becomes significantly weakened or disappears because $k_t^2$ is suppressed to nearly zero for the thickness shear excitation in 128$^\circ$Y-cut LN. The corresponding Bode $Q$ responses in Fig.~\ref{fig6} show the same angular dependence, decreasing as the propagation angle approaches 90$^\circ$. The responses from 110$^\circ$ to 170$^\circ$ are not shown in Figs.~\ref{fig5} and \ref{fig6} because the thickness shear excitation in 128$^\circ$Y-cut LN is expected to be approximately symmetric with respect to 90$^\circ$, as also observed in the FEA results in Fig.~\ref{fig2}. Therefore, the 10$^\circ$--90$^\circ$ range is sufficient to show the monotonic reduction of $k_t^2$ and $Q$ toward the symmetry point.

\begin{figure}[t!]
\centerline{\includegraphics[width=\columnwidth]{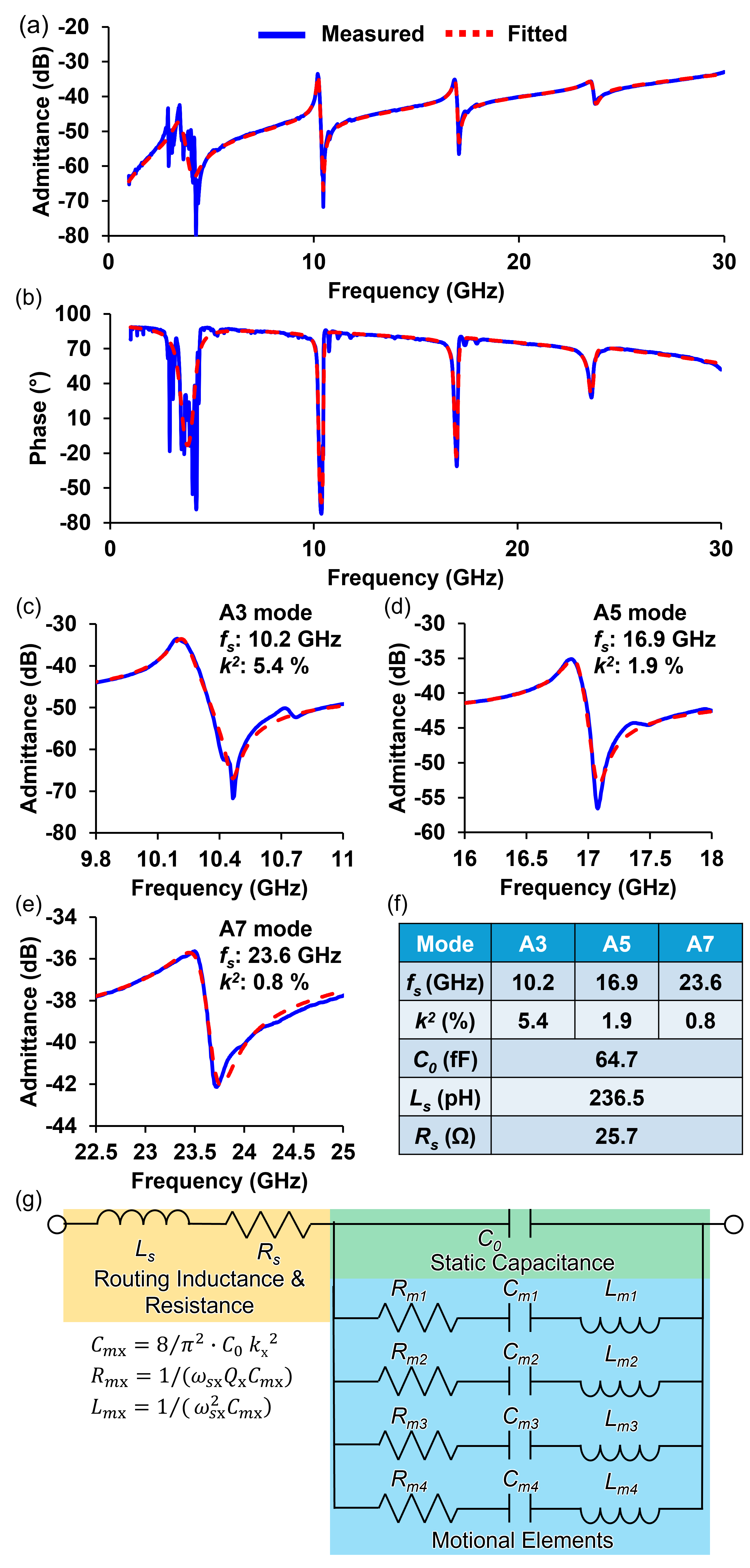}}
\caption{mBVD fitting results and extracted model parameters for the Design A resonator at an in-plane angle of 10$^\circ$: (a) measured and fitted wideband admittance responses from 1 to 30 GHz; (b) measured and fitted wideband phase responses; enlarged admittance responses and mBVD fitting results of the (c) A$_3$, (d) A$_5$, and (e) A$_7$ modes; (f) extracted mBVD model parameters; and (g) equivalent mBVD circuit model.}
\label{fig7}
\end{figure}

To obtain more reliable $k^2$ values from the measured responses, a multi-branch modified mBVD model was used, as shown in Fig.~\ref{fig7}. Although the $f_s$/$f_p$-based method in (1) provides a useful estimate of coupling, it can become unreliable for lossy or low-$Q$ resonators, where the resonance and antiresonance peaks are broadened or shifted \cite{naikElectromechanicalCouplingConstant1998}. In addition, nearby spurious modes can distort the intended resonance response and lead to inaccurate coupling extraction \cite{luAccurateExtractionLarge2019}. At high frequencies, routing inductance, routing resistance, pad capacitance, and other electromagnetic parasitics can further affect the measured admittance response. Therefore, the mBVD model was employed to separate the static capacitance, routing parasitics, and motional branches of the intended resonances, enabling more accurate extraction of $k^2$ from the measured data \cite{larsonModifiedButterworthVanDyke2000}. The representative mBVD fitting results in Fig.~\mbox{\ref{fig7}} were obtained from the Design A resonator at an in-plane angle of 10$^\circ$.
Fig.~\ref{fig7}(a) and (b) show that the measured wideband admittance and phase responses are well captured by the fitted model from 1 to 30 GHz. The enlarged fitting results in Fig.~\ref{fig7}(c)--(e) further confirm that the A$_3$, A$_5$, and A$_7$ resonances are accurately modeled. The extracted model parameters are summarized in Fig.~\ref{fig7}(f), and the corresponding equivalent mBVD circuit is shown in Fig.~\ref{fig7}(g). The $k^2$ values reported in the following statistical analysis are therefore extracted from the mBVD-fitted parameters.

\begin{figure}[t!]
\centerline{\includegraphics[width=\columnwidth]{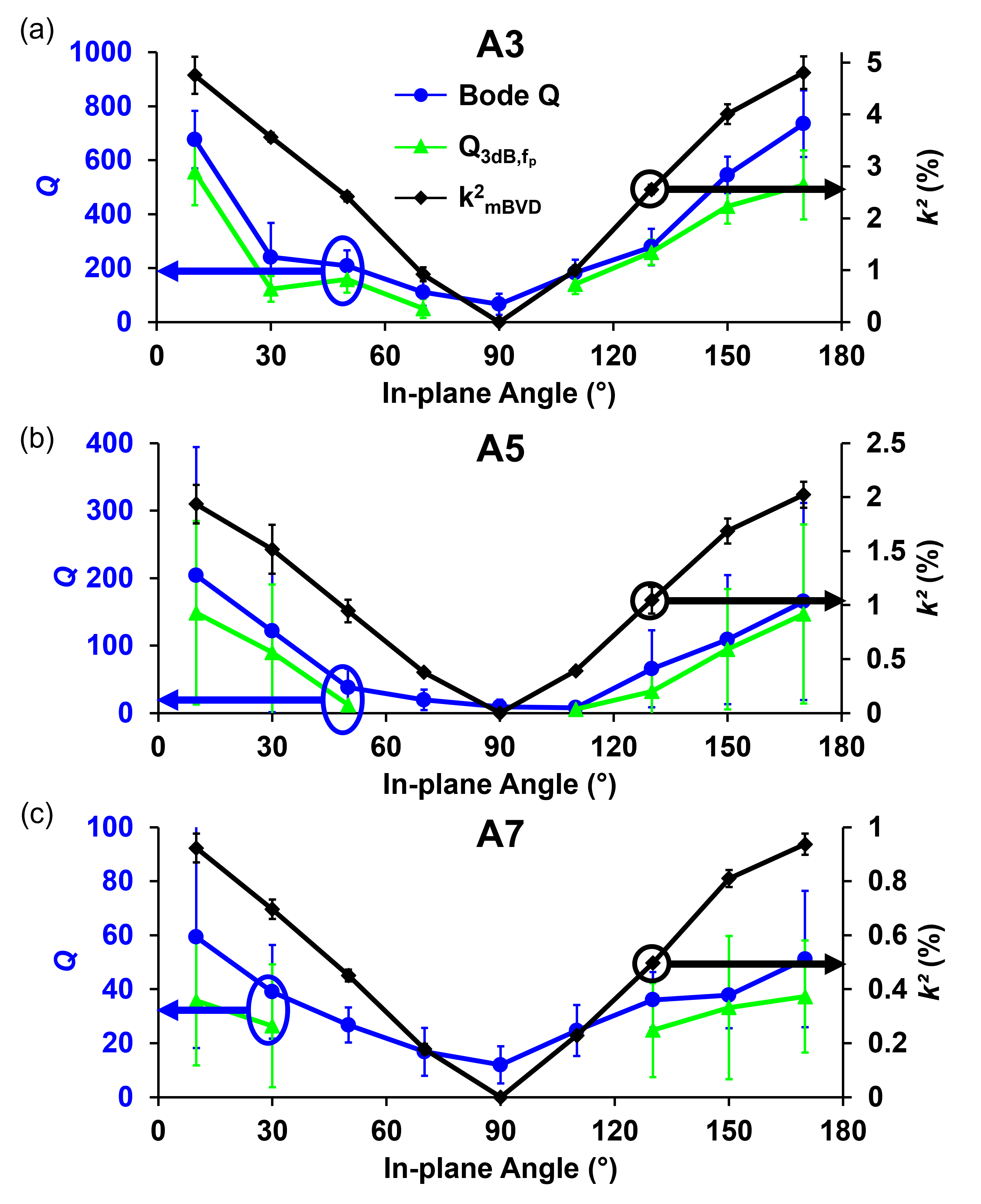}}
\caption{Extracted Bode $Q$, $Q_{\mathrm{3dB}, f_p}$, and mBVD-fitted $k^2$ as functions of in-plane angle for the (a) A$_3$, (b) A$_5$, and (c) A$_7$ modes.}
\label{fig8}
\end{figure}

Figure~\mbox{\ref{fig8}} summarizes the averaged Bode $Q$, $Q_{\mathrm{3dB},f_p}$, and mBVD-fitted $k^2$ values extracted from the four resonator designs as functions of in-plane angle for the A$_3$, A$_5$, and A$_7$ modes. The corresponding individual-design results are provided in Supplementary Figs.~S1--S4. The individual-design data confirm that the overall concave-up angular dependence of $Q$ is observed in each design, following the same general trend as $k^2$. The relatively large error bars, particularly for the A$_5$ and A$_7$ modes, mainly reflect design-to-design differences in the absolute $Q$ values, as discussed in Supplementary Section~S1.
 For each angle, the symbols represent the average values extracted from four resonator designs, and the error bars indicate the standard deviation ($\sigma$). The $Q_{\mathrm{3dB}, f_p}$ values were extracted only when a clear 3 dB bandwidth around the parallel resonant frequency could be identified; otherwise, the data points were left blank. The detailed averaged values for the A$_3$, A$_5$, and A$_7$ modes are summarized in Tables~\mbox{\ref{tab:A3_avg}}--\mbox{\ref{tab:A7_avg}}, and the corresponding individual-design values are provided in Supplementary Tables~S1--S3. The averaged Bode $Q$/$Q_{\mathrm{3dB}, f_p}$ values for the A$_3$/A$_5$/A$_7$ modes are 735/546, 204/149, and 59/37, respectively. At 90$^\circ$, the averaged Bode $Q$ values decrease to 66/9/12, while $Q_{\mathrm{3dB}, f_p}$ cannot be extracted. The averaged $k^2$ values follow the same trend, decreasing from 4.8\%/2.0\%/0.9\% for the A$_3$/A$_5$/A$_7$ modes to nearly zero at 90$^\circ$. Overall, Fig.~\ref{fig8} reveals a consistent concave-up angular dependence in both $k^2$ and $Q$ across all three modes, indicating a strong correlation between anisotropic $Q$ and anisotropic $k^2$ in 128$^\circ$Y-cut LN resonators. Although the resonance response becomes increasingly weak near 90$^\circ$ as $k^2$ approaches zero, making quantitative $Q$ extraction less reliable, the overall agreement in the angular trends of $k^2$ and $Q$ remains evident.
\vspace{1em}

\begin{table}[t!]
\caption{Averaged resonator performance for A$_3$ mode}
\label{tab:A3_avg}
\centering
\begin{tabular}{c c c c}
\hline\hline
Angle ($^\circ$) & $k^2$ (\%) & Bode $Q$ & $Q_{\mathrm{3dB},f_p}$ \\
\hline
10  & 4.8 & 676 & 546 \\
30  & 3.6 & 241 & 123 \\
50  & 2.4 & 208 & 159 \\
70  & 0.9 & 111 & 51  \\
90  & 0.0 & 66  & --  \\
110 & 1.0 & 183 & 140 \\
130 & 2.6 & 278 & 259 \\
150 & 4.0 & 545 & 429 \\
170 & 4.8 & 735 & 509 \\
\hline\hline
\end{tabular}
\end{table}

\begin{table}[t!]
\centering
\caption{Averaged resonator performance for A$_5$ mode}
\label{tab:A5_avg}
\begin{tabular}{c c c c}
\hline\hline
Angle ($^\circ$) & $k^2$ (\%) & Bode $Q$ & $Q_{\mathrm{3dB},f_p}$ \\
\hline
10  & 1.9 & 204 & 149 \\
30  & 1.5 & 122 & 90  \\
50  & 0.9 & 38  & 12  \\
70  & 0.4 & 20  & --  \\
90  & 0.0 & 9   & --  \\
110 & 0.4 & 8   & 6   \\
130 & 1.0 & 66  & 32  \\
150 & 1.7 & 109 & 95  \\
170 & 2.0 & 165 & 147 \\
\hline\hline
\end{tabular}
\end{table}

\begin{table}[t!]
\centering
\caption{Averaged resonator performance for A$_7$ mode}
\label{tab:A7_avg}
\begin{tabular}{c c c c}
\hline\hline
Angle ($^\circ$) & $k^2$ (\%) & Bode $Q$ & $Q_{\mathrm{3dB},f_p}$ \\
\hline
10  & 0.9 & 59 & 36 \\
30  & 0.7 & 39 & 26 \\
50  & 0.5 & 27 & -- \\
70  & 0.2 & 17 & -- \\
90  & 0.0 & 12 & -- \\
110 & 0.2 & 25 & -- \\
130 & 0.5 & 36 & 25 \\
150 & 0.8 & 38 & 33 \\
170 & 0.9 & 51 & 37 \\
\hline\hline
\end{tabular}
\end{table}

\section{Discussion}

\begin{figure}[t!]
\centerline{\includegraphics[width=\columnwidth]{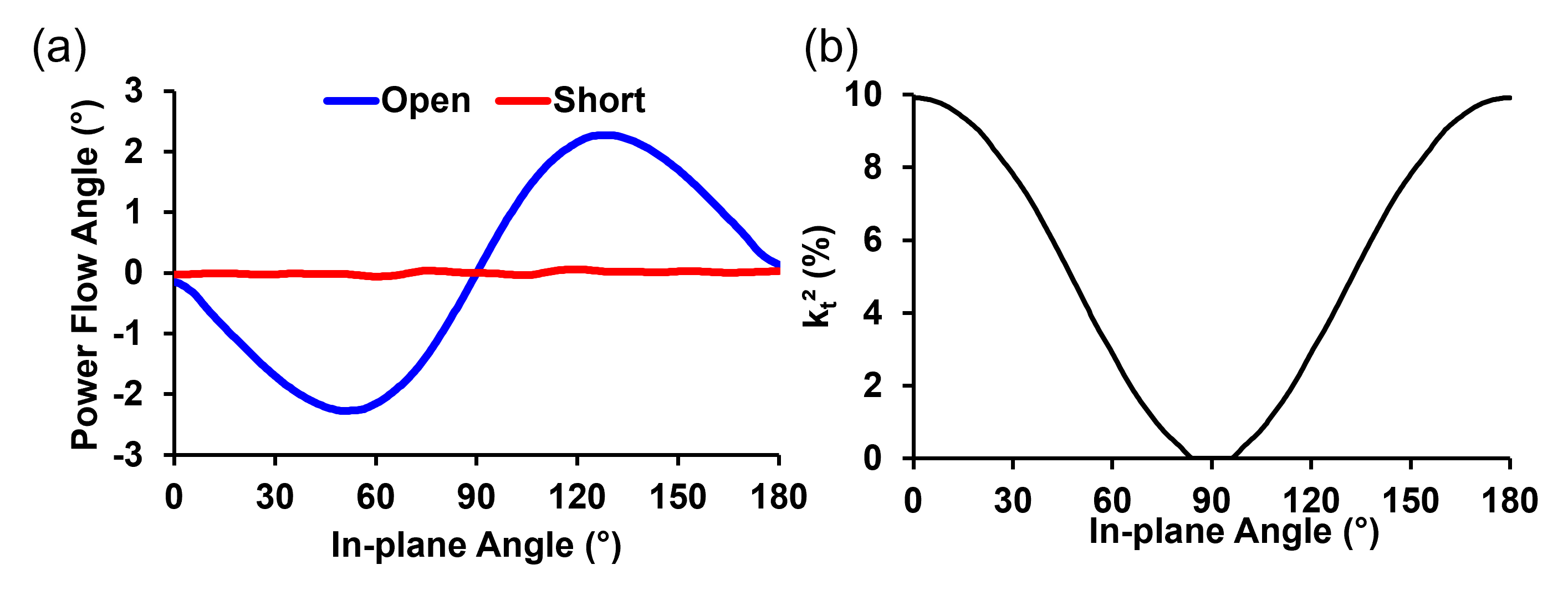}}
\caption{Simulated propagation characteristics of the A$_3$ mode in 128$^\circ$Y-cut LN: (a) power flow angle and (b) electromechanical coupling coefficient $k_t^2$ as a function of in-plane angle.}
\label{fig9}

\end{figure}

The measured results show that the anisotropic $Q$ of the A$_3$, A$_5$, and A$_7$ modes follows a trend similar to $k^2$, with degraded $Q$ near 90$^\circ$. To examine whether this behavior can be explained by propagation-related loss, we considered the power flow angle (PFA) of the A$_3$ mode in 128$^\circ$Y-cut LN. In anisotropic piezoelectric materials, a nonzero PFA can cause the acoustic energy flow to deviate from the intended propagation direction, which may introduce additional propagation loss in resonator devices~\cite{luGigahertzLowLossWideband2019,kuznetsovaPowerFlowAngle2008,luPowerFlowAngles2021,leeVisualizationAcousticPower2021}.

Fig.~\mbox{\ref{fig9}}(a) shows the calculated PFA of the A$_3$ mode as a function of in-plane angle. The electrically short condition shows a relatively small PFA over the full angular range, while the electrically open condition exhibits larger PFA values. The PFA becomes zero at 0$^\circ$, 90$^\circ$, and 180$^\circ$, and its absolute value reaches larger values around 50$^\circ$ and 130$^\circ$. Fig.~\mbox{\ref{fig9}}(b) shows the simulated electromechanical coupling coefficient $k_t^2$, which reaches its maximum near 0$^\circ$ and 180$^\circ$ and decreases toward 90$^\circ$, consistent with the admittance simulation and measurement trends discussed earlier.

However, the measured $Q$ trend cannot be explained by PFA alone. If PFA were the dominant loss mechanism, the largest loss would be expected near the angles where the absolute PFA is maximized, such as around 50$^\circ$ and 130$^\circ$. In contrast, the measured Bode $Q$ and $Q_{\mathrm{3dB},f_p}$ are lowest near 90$^\circ$, where the PFA is close to zero. This discrepancy indicates that, although PFA can contribute to propagation-related loss, it is unlikely to be the dominant cause of the anisotropic $Q$ degradation observed in the 128$^\circ$Y-cut LN resonators. Therefore, another orientation-dependent deformation mechanism must be considered.

\begin{figure}[t!]
\centerline{\includegraphics[width=\columnwidth]{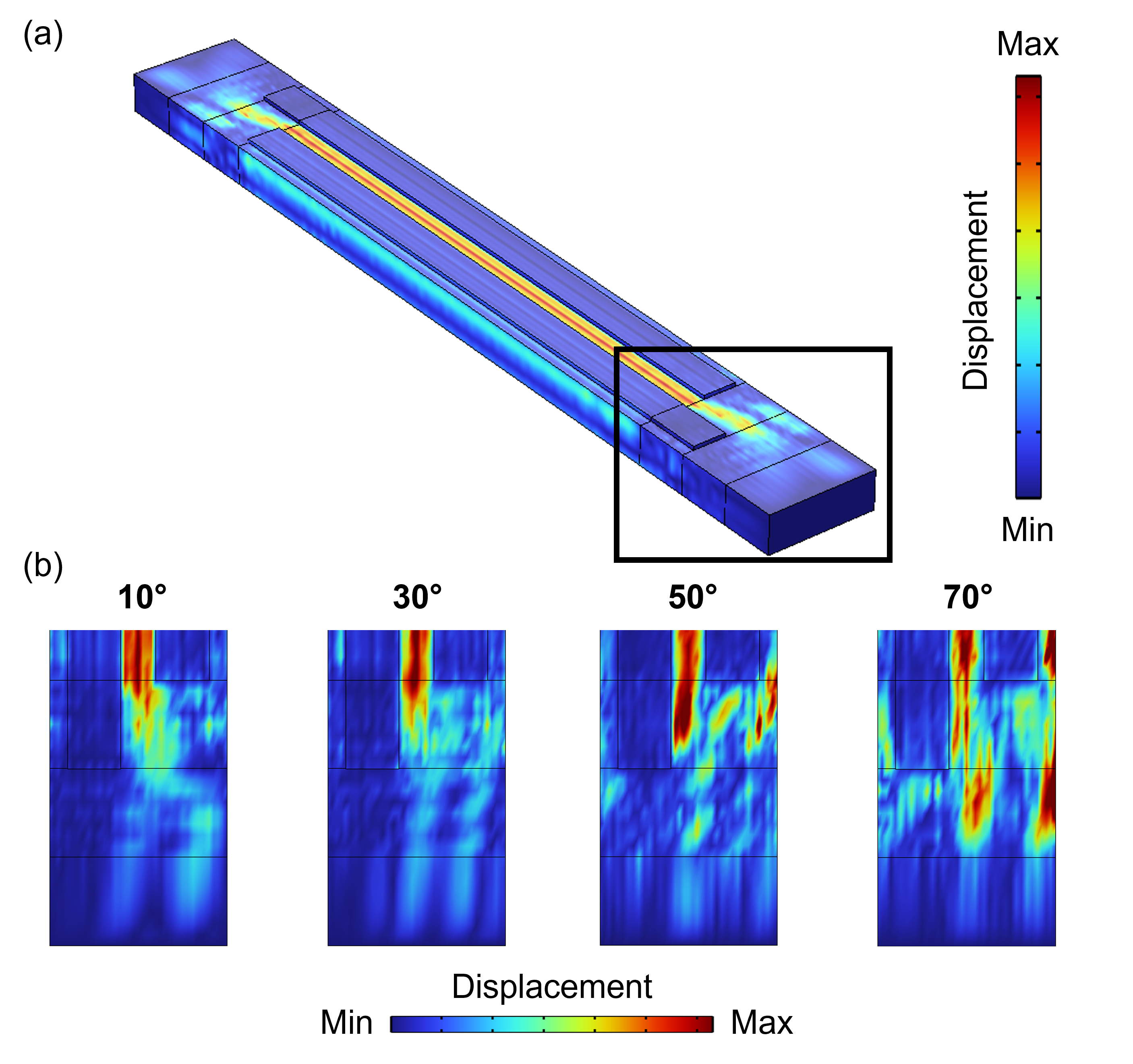}}
\caption{Simulated displacement distributions of the A$_3$ mode evaluated at the parallel resonance frequency ($f_p$): (a) overall displacement profile of a one-pair-IDT resonator; (b) enlarged views of the anchor region at in-plane angles of 10$^\circ$, 30$^\circ$, 50$^\circ$, and 70$^\circ$, showing enhanced displacement leakage toward the anchor as the in-plane angle approaches 90$^\circ$.}
\label{fig10}

\end{figure}

To further investigate this mechanism, we performed full-device FEA using a one-pair-IDT resonator geometry, as shown in Fig.~\mbox{\ref{fig10}}. The displacement profiles in Fig.~\mbox{\ref{fig10}} were evaluated at the parallel resonance frequency ($f_p$) of the A$_3$ mode for each corresponding in-plane angle, namely 10.86, 10.82, 10.7, and 10.58 GHz for 10$^\circ$, 30$^\circ$, 50$^\circ$, and 70$^\circ$, respectively. This $f_p$-based evaluation accounts for the orientation-dependent frequency shift and is consistent with the extraction of $Q_{\mathrm{3dB},f_p}$. This model was used to examine how the displacement field evolves near the anchor region as the in-plane propagation angle changes. As shown in Fig.~\mbox{\ref{fig10}}(b), the displacement component transverse to the intended acoustic propagation direction becomes increasingly pronounced near the anchors as the propagation angle approaches 90$^\circ$, suggesting that the mode develops stronger unintended in-plane transverse motion. In 128$^\circ$Y-cut LN, the effective thickness shear excitation becomes weaker as the propagation angle approaches 90$^\circ$, as indicated by the reduction in $k^2$. Under this condition, a larger portion of the vibration is coupled into an unintended transverse displacement component, which can more easily leak energy through the anchor regions. As a result, the measured $Q$ decreases together with $k^2$, even though the PFA does not predict maximum loss at 90$^\circ$. Therefore, the anisotropic $Q$ behavior in these 128$^\circ$Y-cut LN resonators is more strongly associated with orientation-dependent mode deformation and anchor-region energy leakage than with PFA alone.

Within the loss framework of (2), the measured $Q$ trends can be separated into a mode-dependent baseline and an orientation-dependent excess loss. The reduction in the overall $Q$ level from the A$_3$ to the A$_7$ mode is consistent with intrinsic damping mechanisms that become more pronounced at higher frequencies, including phonon-interaction-related loss \cite{tabrizianEffectPhononInteractions2009,chandorkarLimitsQualityFactor2008}. These intrinsic mechanisms can set the baseline $Q$ for each mode and are not expected to vary strongly with the in-plane propagation angle. By contrast, the angular variation of $Q$ within a given mode is more directly related to the $Q_{\mathrm{anchor}}$. At high-coupling orientations, the intended thickness-shear motion is dominant and the unintended transverse displacement near the anchors remains weak, so the measured $Q$ is closer to its intrinsic baseline. As the propagation direction is rotated toward 90$^\circ$, the weakened thickness-shear excitation is accompanied by stronger transverse displacement near the support regions, increasing anchor-mediated energy leakage and reducing $Q_{\mathrm{anchor}}$. Therefore, the present results should not be interpreted as identifying the dominant loss mechanism at the highest-$Q$ orientation. Rather, they show that the orientation-dependent degradation of $Q$ away from the high-coupling directions is primarily associated with anchor-region energy leakage.

\section{Conclusion}

This work experimentally investigates the anisotropic $Q$ of XBARs in 500-nm-thick 128$^\circ$Y-cut LN resonators on Si. By characterizing A$_3$, A$_5$, and A$_7$ modes from 10$^\circ$ to 170$^\circ$ with respect to the material $x$-axis, clear orientation-dependent trends in both $k^2$ and $Q$ were observed. The averaged Bode $Q$/$Q_{\mathrm{3dB}, f_p}$ values were 735/546, 204/149, and 59/37 for the A$_3$, A$_5$, and A$_7$ modes, respectively, while the average Bode $Q$ values at 90$^\circ$ decreased to 66/9/12. The extracted $k^2$ values showed the same angular dependence, decreasing from averaged values of 4.8\%, 2.0\%, and 0.9\% for the A$_3$, A$_5$, and A$_7$ modes to 0\% at 90$^\circ$. FEA results further reveal that the degraded $Q$ near 90$^\circ$ cannot be fully explained by power flow angle alone. Instead, as the propagation angle approaches 90$^\circ$, the reduced thickness-shear excitation is accompanied by stronger unintended transverse displacement near the anchors, thereby enhancing acoustic energy leakage and lowering $Q$. These results indicate that the orientation-dependent degradation of $Q$ in 128$^\circ$Y-cut LN resonators, rather than the absolute $Q$ level, is strongly linked to orientation-dependent mode deformation and anchor-region energy leakage. This study provides experimental evidence and physical insight into anisotropic loss mechanisms in 128$^\circ$Y-cut LN XBARs, offering design guidance for jointly optimizing $k^2$, $Q$, and propagation orientation in high-frequency acoustic devices.

\section*{Acknowledgment}
This work was completed at the Texas Nanofabrication Facility Microelectronics Research Center.

\bibliographystyle{IEEEtran}
\bibliography{references}

\end{document}